\documentclass{emulateapj}

\usepackage{natbib}

\newcommand{\eg}{e.g.}
\newcommand{\etal}{et al.}
\newcommand{\mic}{\mathrm{\mu s}}

\shorttitle{Giant pulses from PSR~B1937+21}
\shortauthors{Soglasnov \etal}

\begin{document}

\title{Giant Pulses from PSR~B1937+21 with widths $\leq 15$~nanoseconds
and  $T_\mathrm{b} \geq5\times 10^{39}$~K,
the Highest Brightness Temperature Observed in the Universe}

\author{V.~A.~Soglasnov\altaffilmark{1},
   M.~V.~Popov\altaffilmark{1},
   N.~Bartel\altaffilmark{2},
   W.~Cannon\altaffilmark{2},
   A.~Yu.~Novikov\altaffilmark{2},
   V.~I.~Kondratiev\altaffilmark{1}, and V.~I.~Altunin\altaffilmark{3}}
 
\altaffiltext{1}{Astro Space Center of the Lebedev Physical Institute,
                Profsoyuznaya 84/32, Moscow, 117997 Russia; vsoglasn@asc.rssi.ru, mpopov@asc.rssi.ru, vlad@asc.rssi.ru.}
\altaffiltext{2}{York University, Department of Physics and Astronomy,
                 4700 Keele Street, Toronto, Ontario M3J 1P3 Canada; bartel@yorku.ca, wayne@sgl.sci.yorku.ca, sasha@sgl.sci.yorku.ca.}
\altaffiltext{3}{Jet Propulsion Laboratory,
                 4800 Oak Grove Drive, Pasadena, CA 91109 USA; Valery.I.Altunin@jpl.nasa.gov.}

\begin{abstract}
Giant radio pulses of the millisecond pulsar B1937+21 were recorded with the S2 VLBI system at 1.65~GHz
with NASA/JPL's \mbox{70-m} radio telescope at Tidbinbilla, Australia. These pulses have been observed as strong
as 65\,000~Jy with widths $\leq 15$~ns, corresponding to a brightness temperature $T_\mathrm{b} \geq 5\cdot 10^{39}$~K,
the highest observed in the universe. The vast majority of these pulses occur in a $5.8~\mic$ and $8.2~\mic$ window at
the very trailing edges of the regular main pulse and interpulse profiles, respectively.
Giant pulses occur in general with a single spike. Only in one case out of 309~was the structure clearly more
complex. The cumulative distribution is fit by a power law with index $-1.40 \pm 0.01$ with a low-energy but no
high-energy cutoff.  We estimate that giant pulses occur frequently but are only rarely detected.  When corrected
for the directivity factor, 25~giant pulses are estimated to be generated in one neutron star revolution alone.
The intensities of the giant pulses of the main pulses and interpulses are not
correlated with each other nor with the intensities or energies of the main pulses and interpulses themselves.
Their radiation energy density can exceed 300~times the plasma energy density at the surface of the neutron star and
can even exceed the magnetic field energy density at that surface. We therefore do not think that the generation of
giant pulses is linked to the plasma mechanisms in the magnetosphere. Instead we suggest that it is directly related
to discharges in the polar cap region of the pulsar.
\end{abstract}


\section{INTRODUCTION} \label{intro}

Rapid pulse-to-pulse intensity variations are a common property of pulsar radio emission.
Single pulses are often 10-fold stronger than their average pulse. These intensity variations and the characteristics
of the so-called subpulses have been extensively studied for a long time \citep[see \eg,][]{bartel1980, kramer2003} and can be
explained in the frame of the standard polar cap model \citep[\eg,][]{rankin2000}. The most dramatic events are the
so-called giant pulses. They can be even 1000-fold stronger than the regular single pulses from the pulsar. However giant
pulses are not typical for pulsars. They have been observed in only a small number of pulsars, the Crab pulsar B0531+21
\citep[\eg,][]{hankins2000}, the original millisecond pulsar B1937+21 \citep{wolszczan1984, sallmen1995, cognard1996, kink2000},
the millisecond pulsars B1821$-$24 \citep{romani2001} and B0540$-$69 \citep{johnston2003}. In addition, giant pulses were
recently also reported for B1112+50 \citep{ershov2003} and two milliseconds pulsars B1957+20 and J0218+4232 \citep{joshi2003}.
The giant pulses are rarely observed in each of the pulsars, and are arguably the least understood of any of the pulsar phenomena.

The first giant pulses were observed in the Crab pulsar, shortly after its discovery \citep{staelin1968, heiles1970, staelin1970, gower1972},
and were believed for a long time to be a unique property of this particular pulsar. They have been studied for many years
\citep[\eg,][]{friedman1992, lundgren1994, lundgren1995, moffett1996, sallmen1999, hankins2000}. Their remarkable feature,
beside a huge peak flux density, is a very short duration of at least 2~ns \citep{hankins2003}. They occur as single giant pulses
or in pairs or bunches of several giant pulses, in the main pulse as well as in the interpulse window.

Giant pulses were subsequently found in the millisecond pulsar B1937+21 \citep{wolszczan1984}. \citet{sallmen1995} presented the
first analysis of properties of these giant pulses, using a relatively small set of data. They noted that the giant pulses are
located on the trailing edges of both the main pulse and interpulse. A more extensive study was done by \citet{cognard1996}.
They made observations for 44~minutes with the Arecibo radio telescope at 430~MHz with a time resolution of $2~\mic$ and detected
60~giant pulses with peak flux densities greater than 20~times the peak flux density of the average pulse. They found a remarkable
constancy of the longitude where giant pulses occur. However, interstellar scattering limited the effective resolution and prevented
them from conducting a more detailed analysis. \citet{kink2000} observed giant pulses also with the Arecibo telescope,
at the same frequency for 30~min, and then, non-simultaneously also at higher frequencies, where interstellar
scattering is much smaller, namely at 1420~MHz for 4~h, and at 2380~MHz for 26~min, at a time resolution of $0.38~\mic$. The giant
pulses were essentially unresolved at the highest frequency of 2.3~GHz where the scattering time is less than the time resolution.
Further, they appeared within a narrow window of $\leq 10~\mic$ at the base of the trailing edges of the main pulse and interpulse, where
the regular pulsed emission is nearly absent. Giant pulses in the Crab pulsar and in B1937+21 both have a power-law cumulative
distribution of pulse energy, but with different indices.

Recently, evidence for giant pulses was also reported for the millisecond pulsar B1821$-$24 \citet{romani2001}. The observations
were conducted with the Parkes 64-m radio telescope at a frequency of 1516~MHz with a time resolution of $80~\mic$. In three hours of
observations 16~giant pulses were detected with energies exceeding 50~times the mean pulse energy. The giant pulses were unresolved
both in width and in longitude, appearing in just one resolution bin immediately following the peak of the main pulse.

Finally, the Vela pulsar may harbor giant pulses as well. \citet{johnston2001} found that individual pulses have a very broad
distribution of peak flux densities and can be as strong as 40~times the peak flux density of the average pulse. However, it is not clear
whether such pulses are related to true giant pulses \citep[see also,][]{johnston2003b}.

In this paper we present a detailed analysis of 309~giant pulses in PSR~B1937+21, at 1.65~GHz sampled at intervals of 31.25~ns.
In \S~\ref{observations} we describe the observations, the signal processing, and the data analysis. In \S~\ref{results} we present our
results on the shape and width of giant pulses~(\S~\ref{shape}), their arrival times~(\S~\ref{arrival}), their influence (or lack
of it) on other emission characteristics~(\S~\ref{impact}), and their intensities~(\S~\ref{intens}). In \S~\ref{discussion} we discuss
our results and in \S~\ref{conclusions} present our conclusions.

\section{OBSERVATIONS, SIGNAL PROCESSING AND DATA ANALYSIS}
\label{observations}
\subsection{Observations}

The observations of PSR~B1937+21 were made with the NASA Deep Space Network \mbox{70-m}~DSS43 radio telescope at Tidbinbilla,
Australia on 30~May~1999. The data were recorded continuously for 39~minutes with the S2 VLBI system~\citep{cannon1997, wietfeldt1997}
with 2-bit sampling at the Nyquist rate (32~MHz) in the lower sideband from 1634 to 1650~MHz and the upper sideband from 1650
to 1666~MHz. Left circular polarization was observed in both bands. The observations were made {\it in absentia}, in the same
way VLBI observations are made routinely \citep[see][ for similar observations of microstructure of other pulsars]{popov2002a, popov2002b}.
The tapes were shipped to Toronto and played back through the S2 Tape-to-Computer Interface (S2-TCI) at the Space Geodynamics
Laboratory (SGL) of CRESTech on the campus of York University. The S2-TCI system transfers the baseband-sampled pulsar data to
files on hard disks of a SUN workstation for further processing, which includes data encoding, coherent dedispersion, analysis of
normal pulsed emission, and search for giant pulses.  The final analysis was conducted at the Astro Space Center of the Lebedev
Physical Institute in Moscow.

\subsection{Data Encoding and Dedispersion}

We first encoded and then dedispersed the data using the coherent dispersion removal technique originally developed by
\citet [ see \citet{popov2002a, popov2002b} for details]{hankins1971}. For PSR~B1937+21 the dispersion smearing across the
bandwidth of 16~MHz in each sideband channel is 2.1~ms, greater than the pulsar period of 1.6~ms. We therefore had to encode
and dedisperse the whole record of 39~min. Coherent dispersion removal provided a time resolution
of $\Delta \nu^{-1}$ = 62.5~ns in each sideband or of 31.25~ns when the sidebands were combined.
Then, we split the 39-min record into short blocks of 10~s duration each to facilitate further analysis and processing.
For each of these blocks we computed the root-mean-square~(rms) deviation or $\sigma$, the integrated profile, and the on-pulse
and off-pulse spectra.

\subsection{Search for Giant Pulses} \label{search}

Following these procedures we conducted a search of each 10-s block of data for giant pulses. For our statistical analysis our
intention was to set the threshold for the giant pulse detection as low as possible to obtain as large a sample of giant pulses
as possible while at the same time avoiding a large number of accidental noise events.
The search was conducted without smoothing of the data. The probability,~$P$, of confusion with noise is then given through the
distribution of~${\chi}^2$ with two degrees of freedom by $P(a>x)=e^{-x}$, where the intensities~$x$ are expressed in units of
the rms of the noise fluctuations, or~$\sigma$. The observed noise distribution, obtained from the off-pulse data was found to strictly
follow this law. We selected a threshold for the search of giant pulses of~$21\sigma$ which corresponds to a probability of a noise
outburst with an intensity~$a>21\sigma$ of~$7.6\times 10^{-10}$. With a sampling rate of 32~MHz we expected for each sideband one false
giant pulse every 40~s or 56~false giant pulses during the full length of our observations of 39~min over the total range of longitudes
(whole period).

During the processing of the data of the first 5~min it became clear that, in accordance with earlier results by \citet{kink2000}, all
giant pulses occur in two restricted longitude ranges of the average profile, one close to the main pulse and the other close to the
interpulse and each only~$\sim 10~\mic$ wide (see~\S~\ref{arrival}). Outside these windows the number of events
exceeding our intensity threshold of~$21\sigma$ was just as expected for noise outbursts. No true giant pulse was therefore found outside
these windows.

For the narrow windows themselves, the threshold of~$21\sigma$ gives less than one false giant pulse within these windows over the whole
observing time. To increase our sensitivity for giant pulse detections we therefore used a second, lower, threshold of~$17\sigma$ for the
remaining 34~min and only for these narrow windows.

To filter out the true giant pulses three criteria were used. First, if the event exceeded the threshold of~$17\sigma$ in the primary
sideband (which could be either the upper or lower sideband but in any event was the sideband in which the pulse was strongest) and~$5\sigma$
in the secondary sideband and is delayed according to the dispersion measure of the pulsar, then it was also considered a
true giant pulse. The probability for it to be false is much less than~1 in~34~min and therefore negligible.

Second, if the event exceeded the threshold of~$21\sigma$ in any sideband, then it was considered a true giant pulse.  The probability
for it to be false is less than~1 in~34~min.

Third, if the event exceeded the threshold of~$17\sigma$ in the primary sideband but not the threshold of~$5\sigma$ in the secondary sideband,
then we considered the event to be a true giant pulse only if the event showed at least indications of the characteristic scattering
profile in the primary sideband. To convince ourselves that the ``true giant pulses'' so selected were indeed true giant pulses to a
high degree of probability, we aligned all records in the secondary sidebands by the maxima of the giant pulses
in the primary sidebands and added up all these records. We found that the averaged record of the secondary sidebands
showed an event at the expected longitude with the value of $S/N=12.8$, or~72.5~Jy. With these thresholds, we estimate that the
probability of a mis-identification of a false giant pulse noise event as a true giant pulse and the probability of a mis-identification
of a true giant pulse as a false giant pulse noise event, both to be sufficiently low that no more than 10 such error events are expected
to occur in the 34 minute data set.

With these three criteria we finally selected for our further analysis a statistically representative homogenous sample of 309~giant
pulses, 190~in the main pulse window and 119~in the interpulse window, with a probability of less than~3\% of them being false.

\section{RESULTS} \label{results}
\subsection{Shape and Width of Giant Pulses} \label{shape}

\begin{figure}
\includegraphics[width=0.5\textwidth]{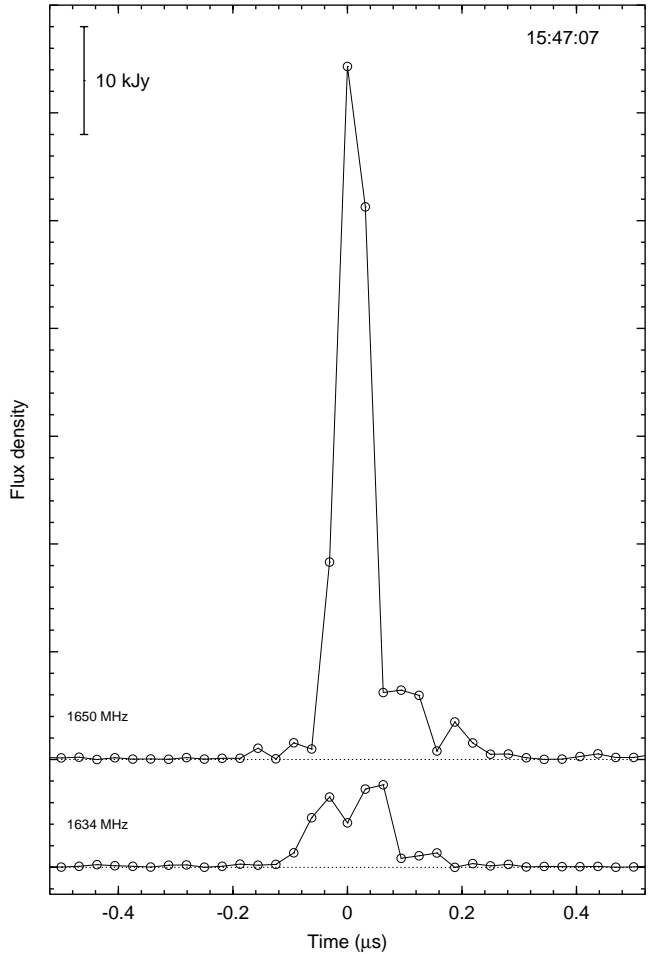}
\caption{
The strongest detected giant pulse at 1634~MHz (lower curve) and at 1650~MHz (upper curve). Dispersion smearing and dispersion delay
are removed. The pulse is shown with the original sampling time of 31.25~ns for each sideband.
\label{fig1}
}
\end{figure}

\begin{figure}
\hskip 8mm\includegraphics[width=0.375\textwidth]{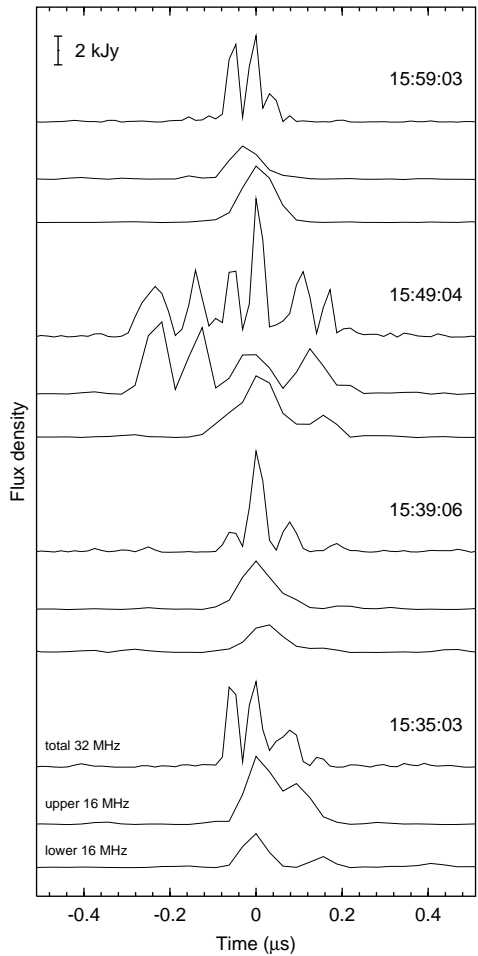}
\caption{
Examples of giant pulses in the lower sideband (lower curve in each set of three plots, 16-MHz band at 1634~MHz), the upper
sideband (middle curve, 16-MHz band at 1650~MHz) and in the two sidebands combined (upper curve, total 32-MHz band). Dispersion
smearing and dispersion delay are removed. The pulses are shown with the original sampling time of 31.25~ns for each sideband
and for 15.625~ns for the total band. The UTC time of their occurrence is given on the right.
The pulses are aligned in phase (in arrival times) by their maxima in the total band.
\label{fig2}
}
\end{figure}

In Figure~\ref{fig1} we display the giant pulse with the highest peak flux density observed during our observations: 65\,000~Jy.
The giant pulse is clearly visible in each of the sidebands but is about sevenfold stronger in the upper sideband.  The width of
the giant pulse in the upper sideband is only about two resolution points or~$\sim 70$~ns.
A few more examples of giant pulses are displayed for the two sidebands (the two lower plots in the sets of three) in Figure~\ref{fig2}.
Like the giant pulse in the previous figure they are mostly all very short. Their width does not exceed a few sample times. Their shape
may be different at the two sidebands, but it always shows an obvious asymmetry: a fast, unresolved, rise time at the leading edge, and
a slower decay at the trailing edge. Such shape is characteristic for interstellar scattering (ISS). We also show a very unusual pulse,
at 15:49:04~UTC. It is the only giant pulse from the total of~309 that shows complex structure, with four spikes of equal amplitude at
1650~MHz and two spikes at 1634~MHz. It may be an example of a giant pulse with intrinsic complex internal structure.

\placefigure{fig1}

For a more detailed analysis of the structure of giant pulses, a portion of the original data, containing 232 giant pulses with sufficient
signal-to-noise ratios was reprocessed with a higher time resolution in the following way. Following data dedispersion,
the spectra for the upper and lower sidebands were concatinated to a single 32-MHz wide band. Then the wide band spectra were inversely
Fourier transformed. The resulting data had now a time resolution twice as short as before. These data were inspected visually and used
for the statistical analysis of the properties of the shape of the giant pulses.

\placefigure{fig2}

Is the decay time of the pulses consistent with the scattering time for B1937+21 at our frequency? The scattering time may be estimated
by extrapolating the scattering times measured at 0.4~GHz \citep{cordes1990b, cognard1996, soglasnov2001, kink2000} and by
analyzing our observed scintillation fringes in the frequency domain (see below).
These procedures lead to a value for the decay time of~$\approx 70$~ns. Therefore, the apparent shape of our giant pulses may be caused by
scattering. Further clues can be obtained from the detailed analysis of the shape, width, and other parameters of the giant pulses. We first
consider the statistics of the rise time and width of the giant pulses.

\subsubsection{The Statistics} \label{stat}

\begin{figure}
\includegraphics[width=0.5\textwidth]{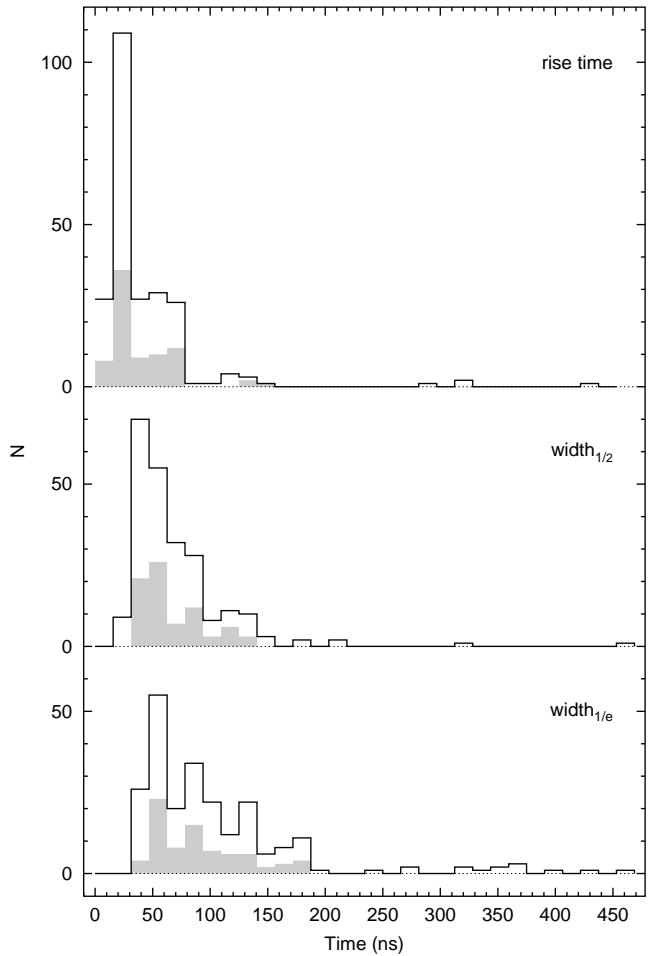}
\caption{
The statistics of the rise time and the width of giant pulses. The rise time is measured from the 1/e~level to the peak of the
pulse (top panel), and the width is measured at a level of~1/2 (middle panel) and~1/e (lower panel) of the maximum flux density.
The black line histograms are for the 232~pulses reprocessed in the total 32-MHz band (see text), and the grey histograms are for
76~strong giant pulses with a peak flux density exceeding~1\,600~Jy.
\label{fig3}
}
\end{figure}

The statistics of the rise time and width of the giant pulses are presented in Figure~\ref{fig3}. We define the rise time as the time
interval between the first sample exceeding~1/e times the maximum flux density and the sample at which the maximum flux density was
measured. The width was defined in two ways. Firstly it is defined as the interval between the first and the last sample for which
the flux density exceeded~1/2 of the maximum and secondly for which it exceeded~1/e of the maximum.

\placefigure{fig3}

\begin{figure*}
\includegraphics[scale=0.53,trim=-25mm 0mm 0mm 0mm]{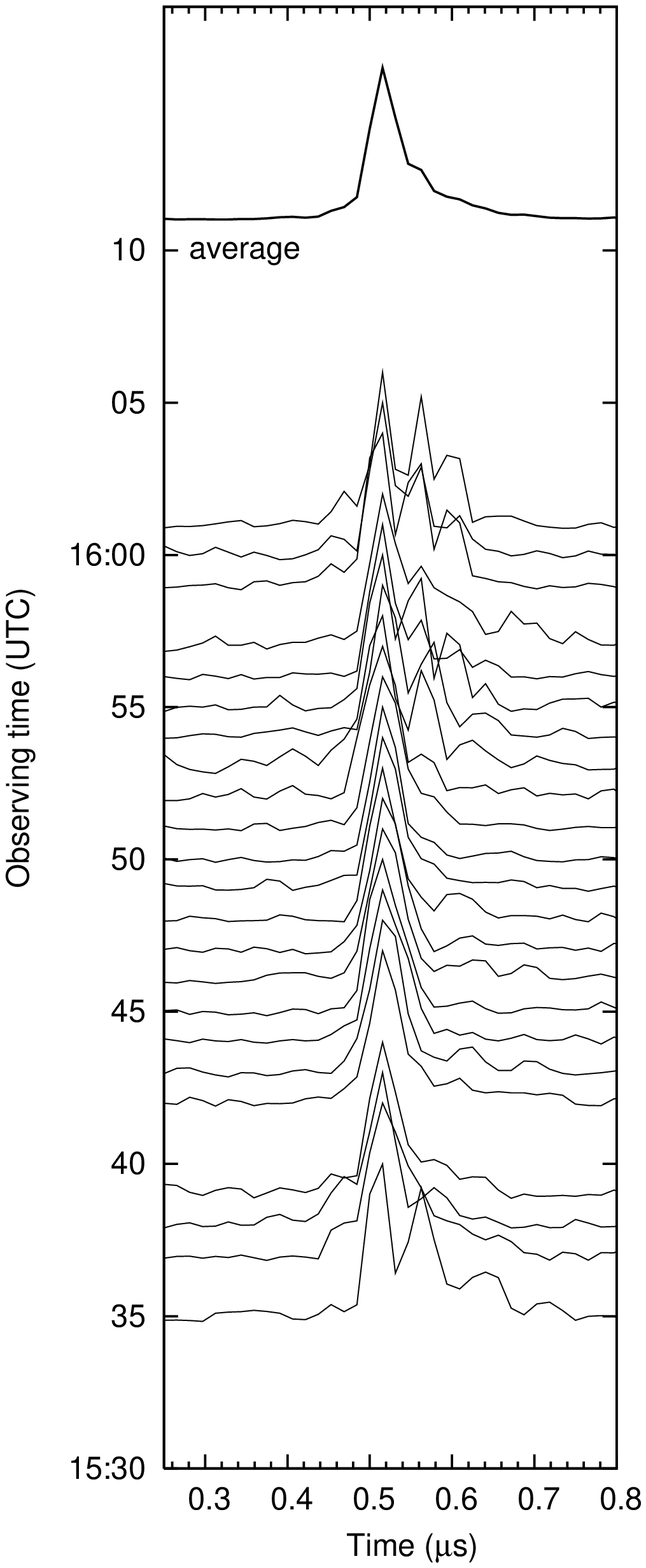}
\includegraphics[scale=0.53,trim=12mm 0mm 0mm 0mm]{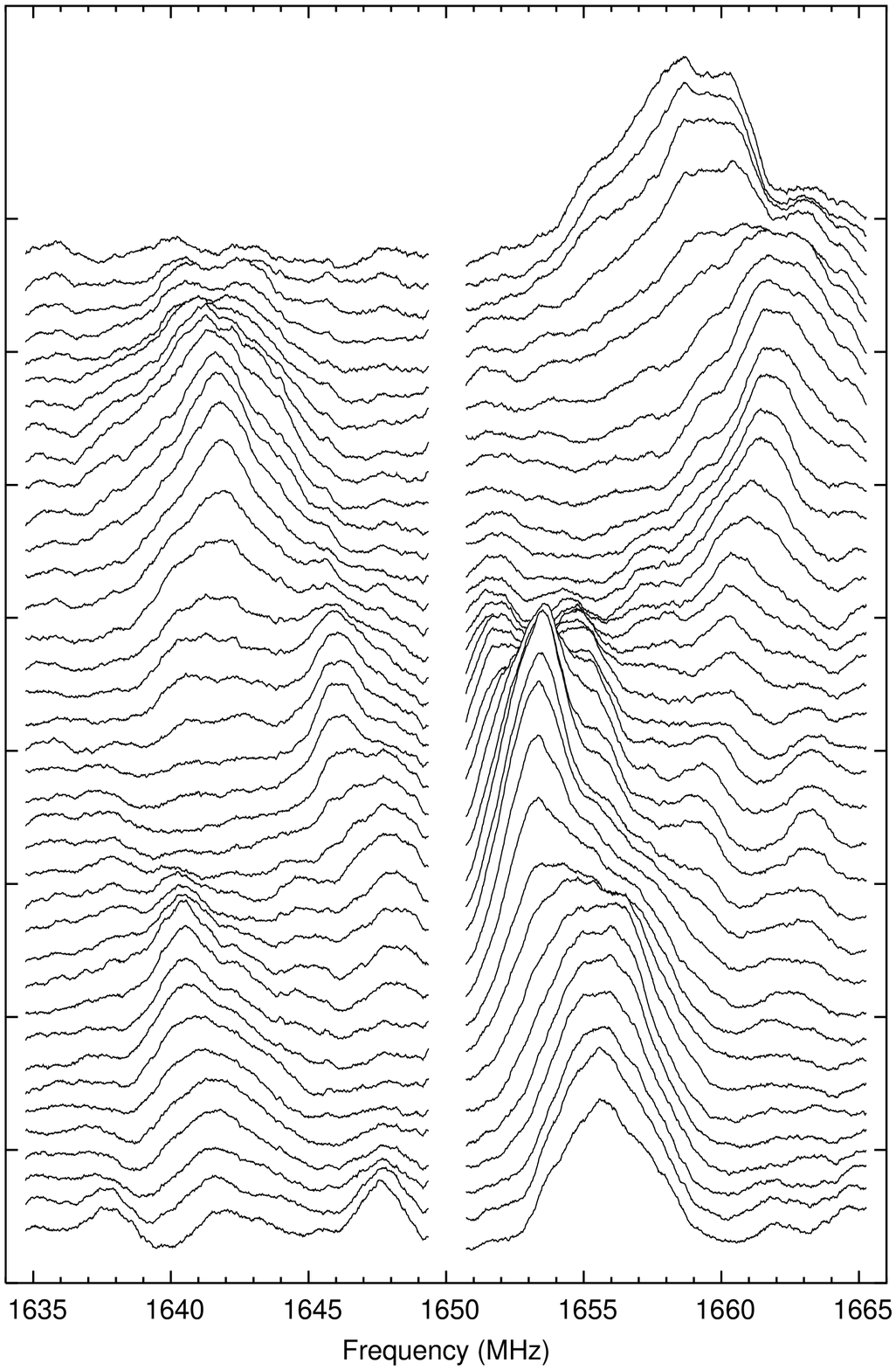}
\caption{
The evolution of the shape of the giant pulses (left panel) and the corresponding dynamic spectra (right panels, for lower and
upper sidebands computed separately) in each case averaged over the same one-minute intervals.
Before averaging, the pulses were normalized by their maxima and aligned by their leading edge at the level of~1/2 of their peak
flux density. If the number of pulses in any given one-minute interval was less than five, the one-minute average pulse was not
plotted. An average giant pulse for the total time of 34~minutes is plotted at the top.
\label{fig4}
}
\end{figure*}  
The distribution of the rise times has a pronounced peak at the 15.6--31.25~ns interval. This corresponds to one sample only.
The leading part of these giant pulses, about half of all our giant pulses, is unresolved. In general, the distribution is quite
narrow. Almost no giant pulses are found with a rise time larger than 75~ns corresponding to 2.5~independent samples.

The distributions of the widths are significantly skewed. The maxima occur at about 40~ns for the 1/2-width and at about 50~ns
for the 1/e-width corresponding in each case to less than two independent samples. The number of giant pulses drops rapidly with
increasing width. Giant pulses of a 1/2-width greater than 150~ns and a 1/e-width greater than 190~ns are absent.

The contribution of noise can be estimated in each of the histograms by comparing them with those of strong giant pulses. If we
restrict the set of giant pulses to only those with a peak flux density exceeding 1\,600~Jy (grey histograms) then it can be seen
that the noise has no significant influence on the main shape of the histograms, except that the long tail visible in each of the
histograms of all giant pulses is completely given by noise.

\subsubsection{The Scatter-Broadened Waveform} \label{wave}

At first sight giant pulses show a large variety of shapes. Nearly half of all giant pulses have a single dominant spike as also
indicated in the histograms discussed in~\S~\ref{stat}. The other half have more complex structure. They contain up to
six strong spikes over a 100--150~ns time interval. In addition, weak emission may accompany the set of spikes just before and
after them.  

However the variations in shape are not random. The shapes evolve slowly over time scales of minutes (see Figure~\ref{fig4}).
Such behavior is typical if caused by interstellar scattering \citep[ISS, \eg,][]{cordes1990b}. Also, the waveform of the average
giant pulse as displayed in the upper part of Figure~\ref{fig4} is well fit by an exponential decay with the characteristic time
of~$65 \pm 5$~ns, very close to the envelope of the scattered pulse predicted in the simple thin screen model. Therefore, it seems
that ISS can account for the variations of the shape of giant pulses and that the true duration of giant pulses is much smaller than
the timescale of the broadening of the pulses by ISS.

\placefigure{fig4}

As a check we compare the evolution of the giant pulse shape with the dynamic spectra of the normal pulsed emission both in the main
pulse and in the interpulse windows. Figure~\ref{fig4} demonstrates a good correspondence between the shape of the giant pulses and
characteristics of the dynamic spectra. In particular, if only one feature dominates the dynamic spectra as at times
\mbox{15:44--15:48~UTC}, the giant pulse is just a single spike. Two or more strong features in the
dynamic spectra are associated with a complex shape of the giant pulse. A change of the features in the dynamic spectra is accompanied
with a rapid change in giant pulse shape. Qualitatively this behavior is in agreement with the well-known correspondence between
the envelope and the power spectrum of a signal.

\begin{figure*}
\mbox{\includegraphics[scale=0.49,trim=-13mm 0mm 0mm 0mm]{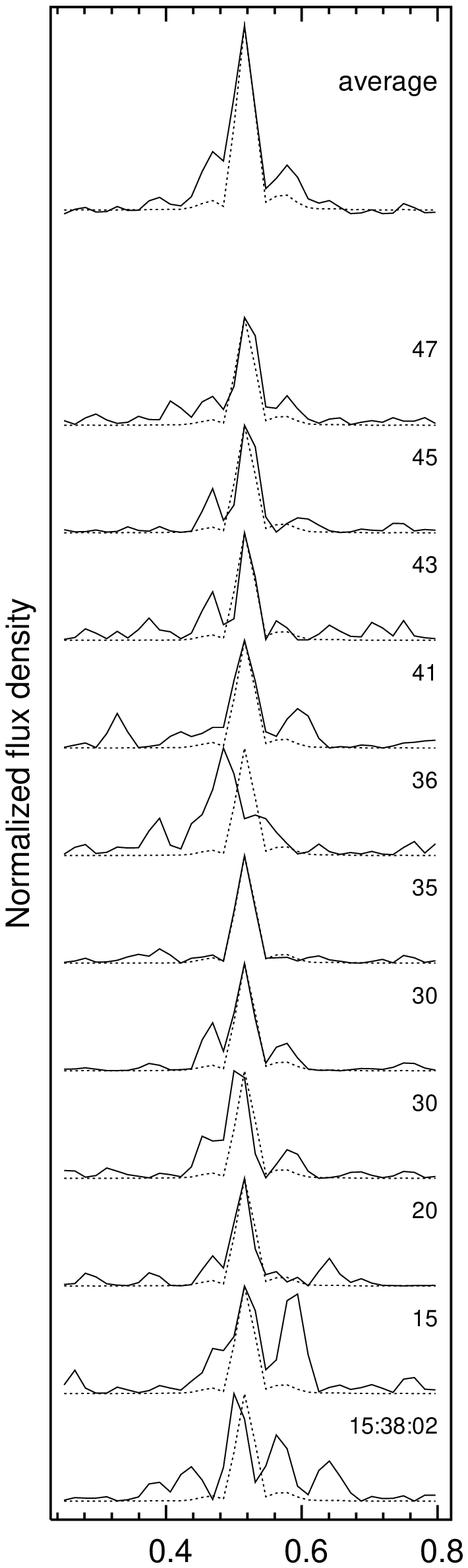}
\includegraphics[scale=0.49,trim=17mm 0mm 0mm 0mm]{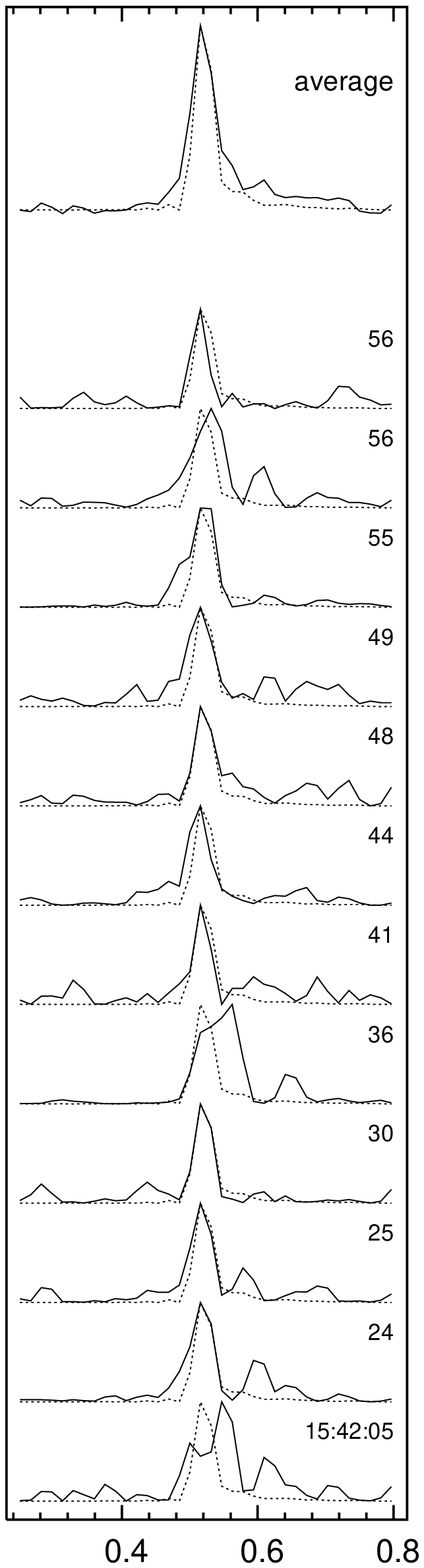}
\includegraphics[scale=0.49,trim=17mm 0mm 0mm 0mm]{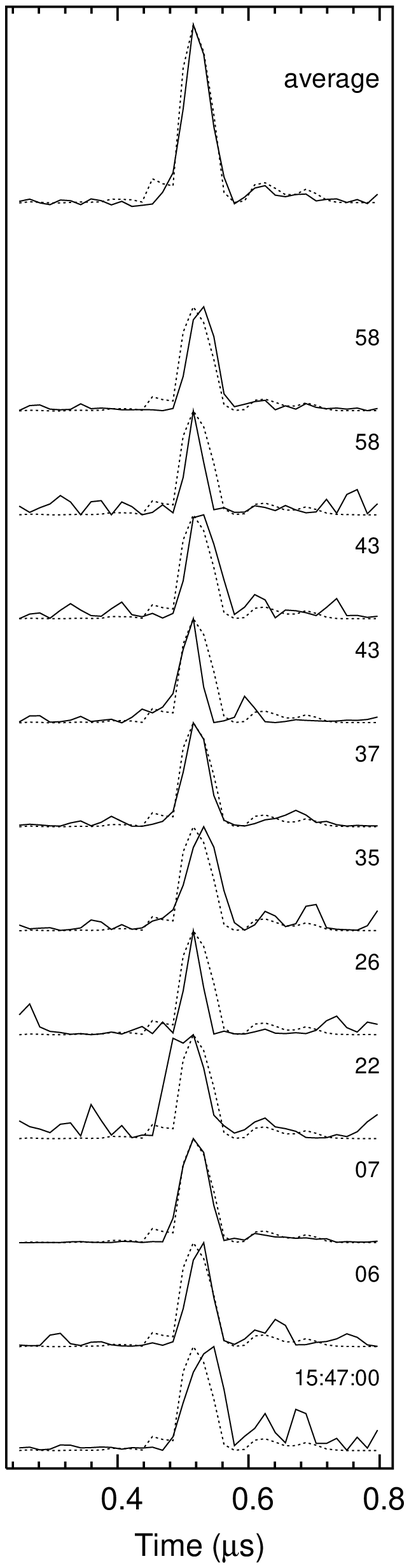}
\includegraphics[scale=0.49,trim=17mm 0mm 0mm 0mm]{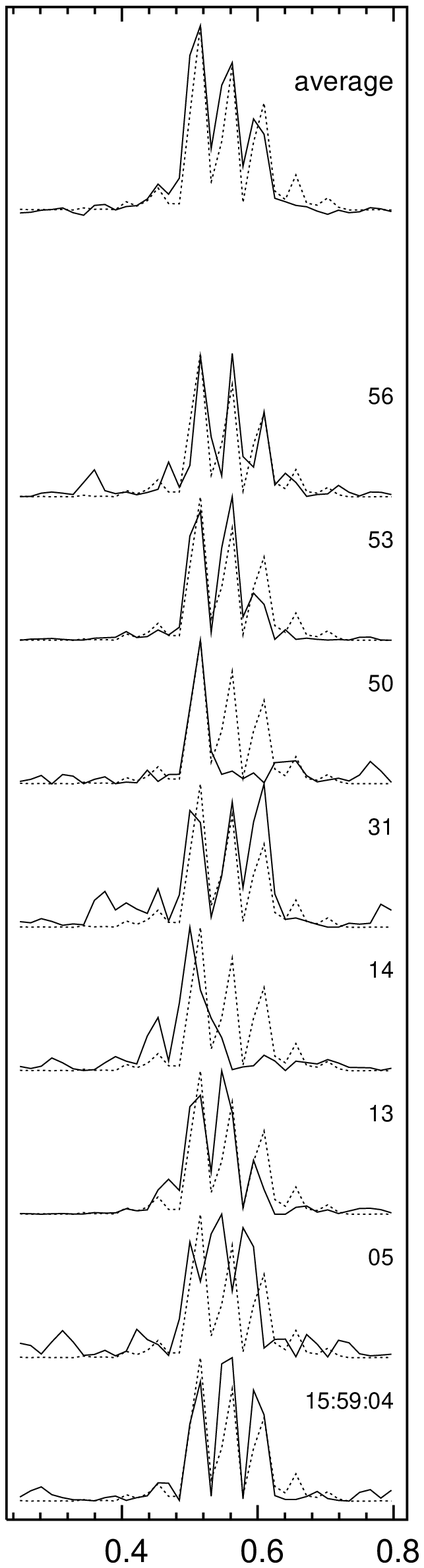}
\includegraphics[scale=0.49,trim=17mm 0mm 0mm 0mm]{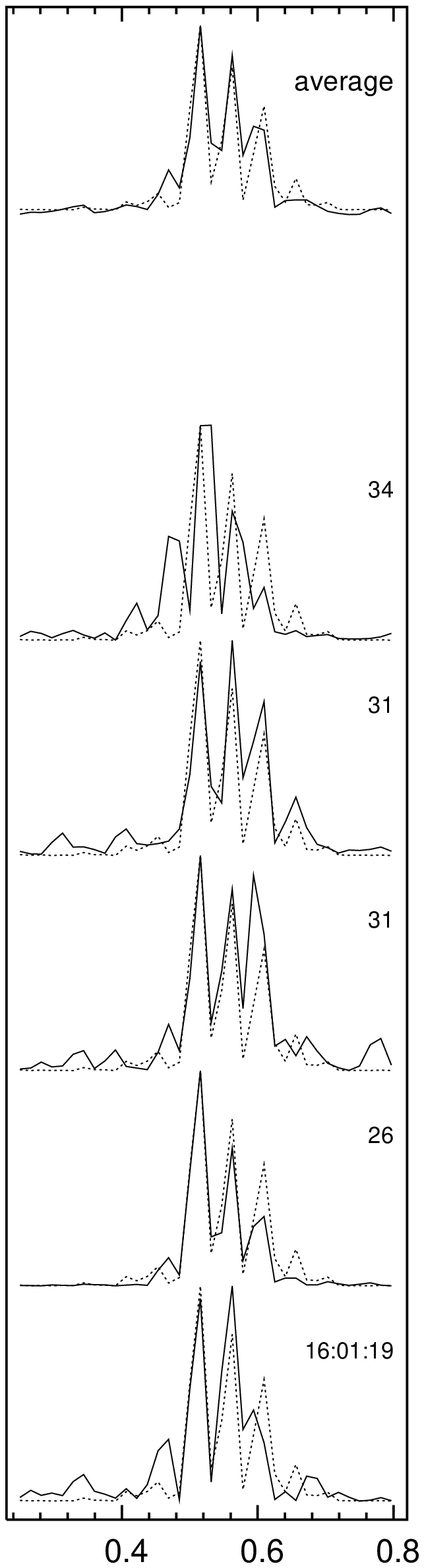}}
\caption{
A collection of giant pulses. Their arrival times in hh:mm:ss are given on the right part of each column with hh:mm remaining the
same in each column. The dotted line accompanying each plot is the waveform of the scattered $\delta-$pulse calculated for the current
minute from the spectra in Figure~\ref{fig4}. The mean profile of the giant pulses in each column is shown at the top. The pulses are
restored for the full 32-MHz band with a time resolution of 31.25~ns and aligned by their leading edge at the level of~1/2 of the
maximum.
\label{fig5}
}
\end{figure*}  

\subsubsection{The Intrinsic Width of Giant Pulses} \label{intrinsicwave}

How short are the giant pulses when they are corrected for the ISS effects?  A full correction for such effects is difficult if not
impossible. However part of the effect can be estimated and a width determined that is less affected by scattering.  
We calculated the response of the scattering medium to a  $\delta$-pulse and later also to broader pulses and compared
the response with the shape of the observed giant pulses.  The impulse response of the scattering medium, $h(t)$,
is given as
\begin{displaymath}
h(t)=\int H(\omega) e^{i\omega t} d\omega \equiv \int
\vert{H(\omega)}\vert e^{i \Phi (\omega)} e^{i \omega t} d \omega~,
\end{displaymath}
where $H(\omega)$ is the complex transfer function of the scattering medium. Both, magnitude $\vert{H(\omega)}\vert$ and
phase $\Phi (\omega)$ may be obtained from the power spectrum of radio emission \citep[for instance, see][ p.~553]{hahn1996, gonor1977}.
In our case, we computed $h(t)$ in each minute of the observing time directly from the dynamic spectra of the normal
emission which are displayed as power spectra in the right panels of Figure~\ref{fig4}.
Then we used a $\delta$-pulse as input to the scattering medium and compared the response minute by minute
with the observed giant pulses averaged minute by minute as displayed in the left panel of Figure~\ref{fig4}.
We found very good agreement. In particular, the number of the spikes and their separations from each other agreed almost exactly.
The only differences were sometimes seen in the relative peak amplitudes. These differences may have been caused by the noise
contribution and by the variations of the scintillation pattern during time intervals shorter than one minute.

Even better agreement was found between the $\delta$-pulse response and individual giant pulses which are plotted in Figure~\ref{fig5}.
In particular, the shape of the spikes agrees almost exactly. The two right columns show a set of giant pulses with three or even more
spikes. Here the similarity is indeed striking. The number of the spikes and their separations from each other agree almost exactly.
This similarity is also clearly visible in the upper parts of the panels where the average giant pulses are displayed.
Such excellent agreement between the responses to the $\delta$-pulse and the observed giant pulses is strong evidence for the width of
giant pulses being much shorter than our time resolution and their apparent shape being due to interstellar scattering.

\placefigure{fig5}

\begin{figure}
\includegraphics[width=0.5\textwidth]{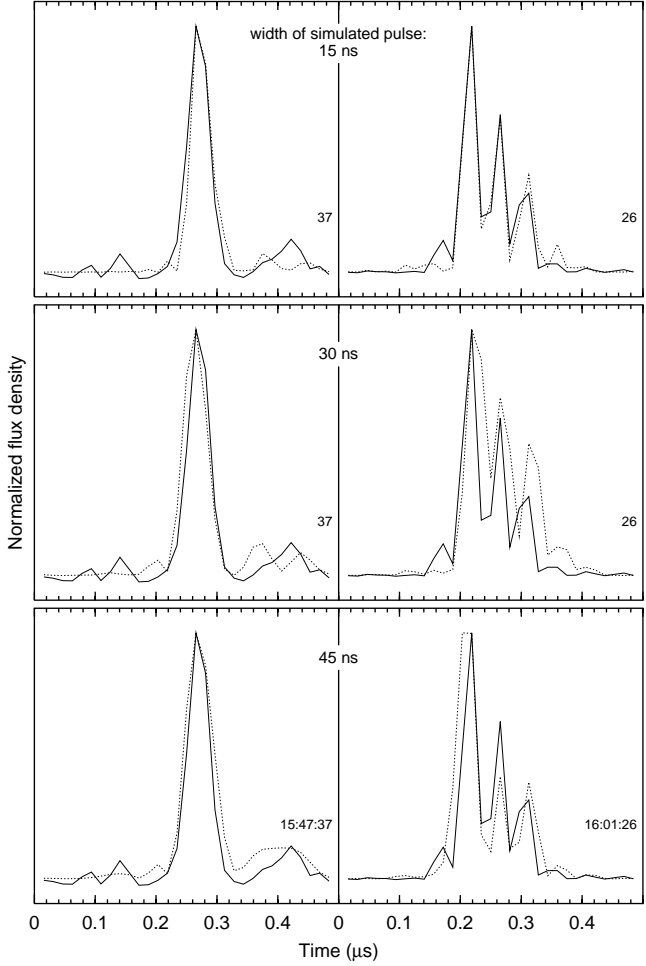}
\caption{
The waveforms simulated in the full 32-MHz synthesized band for two individual giant pulses with different shapes: a single spike
for the giant pulse at UTC~15:47:37 and a triple spike for the giant pulse at UTC~16:01:26. Solid lines show the observed giant pulse,
and the dotted lines show the simulated scattered pulse with different widths of one sample or 15.625~ns (top), two samples or 31.25~ns
(middle), and three samples or 46.875~ns (bottom).
\label{fig6}
}
\end{figure}

To estimate an upper limit of the width of giant pulses we repeated the simulation discussed above, but now with pulses of different widths,
namely of about~15, 30, and 45~ns. In Figure~\ref{fig6} we show the result of the simulation and the comparison with two observed individual
giant pulses, one with a single spike and another one with complex structure. There is no significant difference between the simulated
pulses and the observed giant pulse when the observed giant pulse has just one spike. However, differences become apparent when the
observed giant pulse is complex and the simulated pulse is relatively broad. Only for a simulated pulse with a width of about 15~ns do we
get a good agreement with the observed giant pulse. So, it is obvious that the apparent giant-pulse waveform is a consequence of the
interference of diffraction maxima in the scintillation patterns, and that the true width of giant pulses is equal to, or even smaller
than, 15~ns.

\placefigure{fig6}

\subsection{Giant-Pulse Arrival Times} \label{arrival}

\begin{figure}
\includegraphics[width=0.5\textwidth]{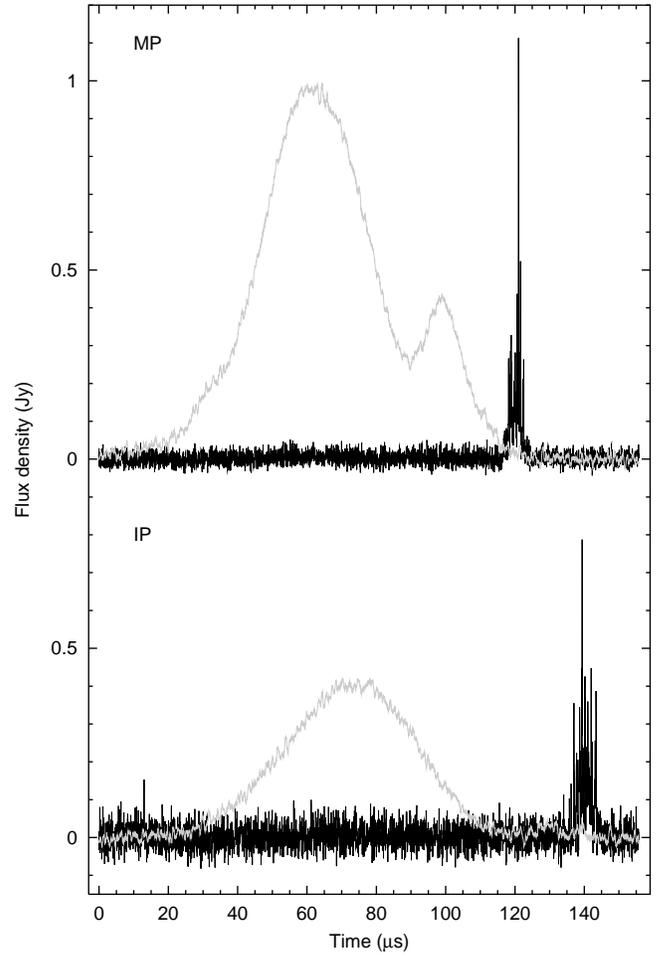}
\caption{
The longitudes of giant pulses in comparison with those of the components of the average profile. The light grey lines show profiles
averaged over the whole observing time of 39~min (1\,501\,344~periods). The black lines represent a profile averaged over only those
periods in which giant pulses were detected, 190~for the main pulse region (MP, upper panel) and 119~for the interpulse region
(IP, bottom). The average profiles are smoothed by a~$0.25~\mic$ running interval while the giant pulse profiles are not smoothed.
\label{fig7}
}
\end{figure}

\begin{figure}
\includegraphics[width=0.5\textwidth]{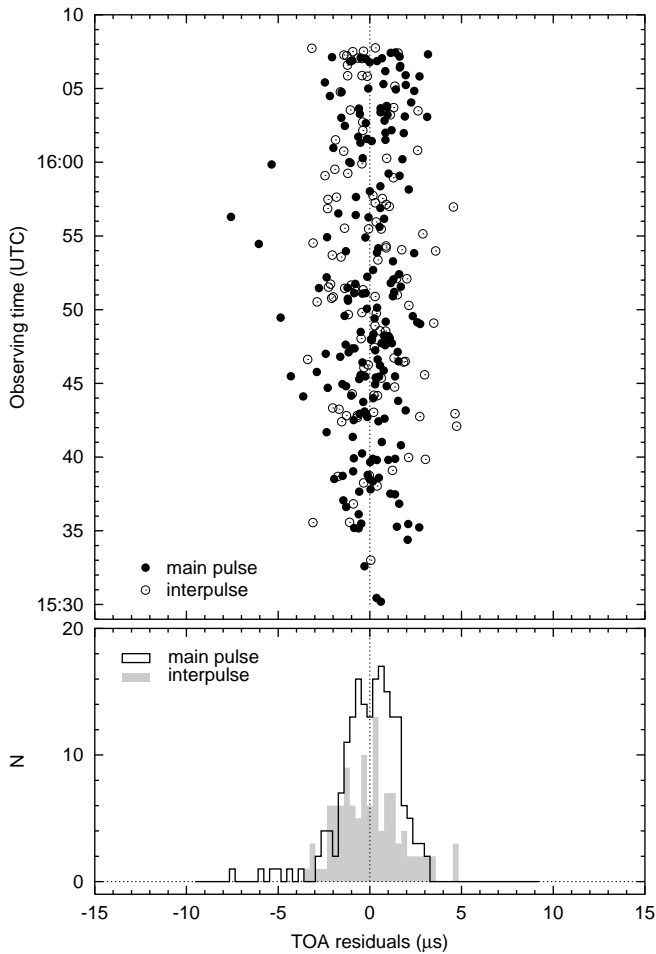}
\caption{
Arrival time residuals for the giant pulses of the main pulse (filled circles) and the giant pulses of the interpulse (open circles).
A residual of~$1~\mic$ corresponds to a residual of~$0\fdg 6$ of longitude. Histograms of residual distributions are shown at the bottom,
a black line for the main pulse and the grey filled area for the interpulse.
\label{fig8}
}
\end{figure}

One of the most striking properties of giant pulses is their occurrence in well-defined, narrow longitude ranges at the trailing edges
of the main pulse and interpulse (Figure~\ref{fig7}). This particular property has been first reported by~\citet{kink2000}.
Figure~\ref{fig8} shows giant pulse arrival time residuals versus observing time and the corresponding histograms.

\placefigure{fig7}

As can be seen from the two figures, the total range of longitudes near the main pulse where giant pulses occur (GPM window) is
$10.7~\mic$, or $2\fdg 5$ in angular units, with an rms pulse-to-pulse arrival time jitter of~$1.6~\mic$ ($0\fdg 37$).
However, the vast majority of giant pulses, namely~97\%, occur within a narrower window of~$5.8~\mic$ with an rms arrival time
jitter of only~$1.2~\mic$ ($1\fdg 3 \pm 0\fdg 3$). The remaining~3\% are 6~pulses
occurring at earlier phases over a longitude interval of~$4.8~\mic$ ($1\fdg 1$). Near the interpulse the width of the corresponding
giant pulse window (GPI window) is~$8.2~\mic$ with an rms arrival time jitter of~$1.7~\mic$ ($1\fdg 9 \pm 0\fdg 4$).  

\placefigure{fig8}

The longitude of giant pulses relative to the normal mean profile is very remarkable. Giant pulses appear at the very trailing edges
of the main pulse and the interpulse where normal emission is almost absent. The center of the main pulse giant pulse  window, defined
as the mean phase of giant pulses, is delayed by~$58.3 \pm 0.3~\mic$ from the peak of the strongest component of the average main pulse
and by~$20.8 \pm0.2~\mic$ from the peak of the second component. The latter value is given more precisely, since this second component
is narrower. For the interpulse region, the center of the GPI window is delayed by~$65.2 \pm 0.5~\mic$ relative to the peak of
the average interpulse (the interpulse profile has a flat top with two small lobes at the edges and one near their midpoint which we
take as a reference point). These measurements are consistent with those of \citet{kink2000} who found delays of~$57~-~58~\mic$
relative to the strongest component of the main pulse and $65~-~66~\mic$ relative to the peak of the interpulse.

When the giant pulses were sufficiently strong in both sidebands we were able to measure the time delay between them at 1650 and 1634~MHz,
and determine the value of the dispersion measure. We obtained $\mathrm{DM}=71.036\pm 0.004~\mathrm{pc}\cdot \mathrm{cm}^{-3}$ where
the uncertainty indicates the range of our DM measurements. Compared with, \eg, those found by \citet{popov2003} ($\mathrm{DM}=71.025$)
and \citet{cognard1995} ($\mathrm{DM}=71.041$) our value lies between these two determinations.

\subsection{Do Giant Pulses Affect Other Emission Characteristics?}
\label{impact}

\begin{figure}
\includegraphics[width=0.5\textwidth]{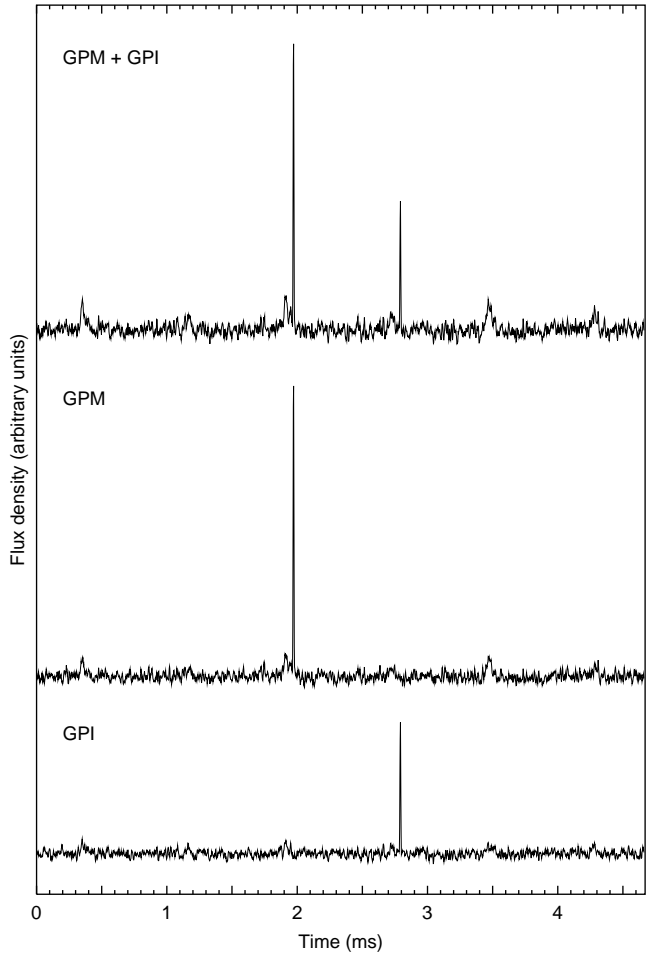}
\caption{
The average profile for all giant pulses (top) and separately for giant pulses in the main pulse (middle) and in the interpulse windows
(bottom), plotted for the period in which the giant pulse occurs together with the periods immediately preceeding and following the event.
\label{fig9}
}
\end{figure}

Is the occurrence of giant pulses correlated with some other parameters of the emission? For instance it may be possible that the energy of the
normal pulse is enhanced or diminished by the occurrence of a giant pulse. In Figure~\ref{fig9} we plot the average profile
for those periods where giant pulses occurred either in the GPM window or the GPI window,
together with one period before the event and one period after the event. In addition we distinguish between main pulse giant pulses and
the interpulse giant pulses and plot the three-period average profiles also for all events where the giant pulses occur in the GPM
window and then where they occur in the GPI window (GPM + GPI).

\placefigure{fig9}

We have not found any clear change in the properties of the normal pulsed emission during giant pulse events. The main pulses and
interpulses keep their usual width and amplitude. For instance, despite the fact that the mean energy of a single giant pulse
is 3 to 5 times larger than the mean energy of a single normal pulse, we found that the occurrence or absence of a giant
pulse changes the mean energy of a single normal pulse by no more than 30\%. Further, the occurrence of a giant pulse in one window
does not trigger the occurrence of a giant pulse in the other window. Giant pulses in the main pulse
and interpulse windows are uncorrelated.

\subsection{Intensity} \label{intens}
\subsubsection{Interstellar Scintillations} \label{scint}

For the study of the intensity variations of giant pulses, the effect of interstellar scintillation on the intensity needs to be taken
into account and corrections need to be applied. As in case of interstellar scattering, we used the normal main pulse and
interpulse emission as reference.

\begin{figure}
\includegraphics[width=0.5\textwidth]{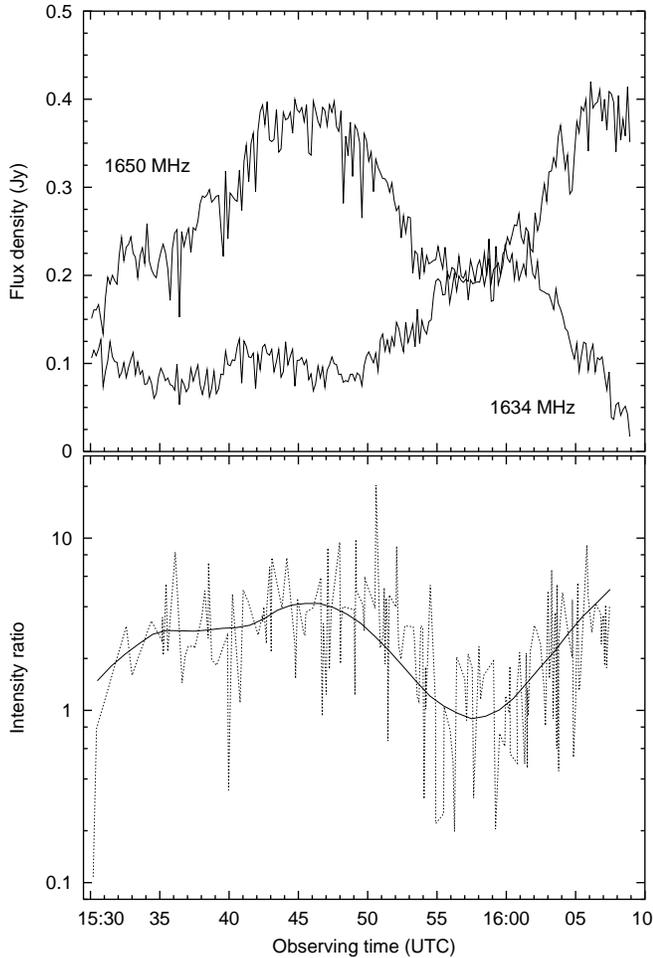}
\caption{
Upper diagram: The scintillation of the intensity of the main pulse in the upper sideband (upper curve, 1650~MHz) and the lower
sideband (lower curve, 1634~MHz) as a function of observing time. Lower diagram: The ratio between the intensities in the upper and
those in the lower sidebands for the normal pulsed emission in the main pulse window (solid line) and for giant pulses (dotted line),
again as a function of observing time.
\label{fig10}
}
\end{figure}

In Figure~\ref{fig10} (upper panel) we show the variation of the intensities of the main pulse in the upper and lower sidebands over
the course of the observations. The intensities vary by more than a factor of two. Such relatively strong variation is expected when
inspecting the characteristics of the dynamic spectra in (Figure~\ref{fig4}). There the ISS diffraction features are small in number,
have lifetimes of only a portion of the observing time, namely 8~to 20~min and widths of only a portion of the bandwidth in each
sideband, namely 2~to 10~MHz. The intensity variations in the two bands as shown in Figure~\ref{fig10} are therefore certainly due to ISS.

\placefigure{fig10}

The variation of the intensities of the giant pulses over the observing time, in contrast, is much stronger. A direct comparison with the
variation of the intensities of the main pulse is not useful. Instead, we computed the ratio between the intensities in the upper and
lower sidebands for a) the giant pulses and b) the main pulse (Figure~\ref{fig10}, lower panel). This ratio is independent of the
initial strength of the emission. As can be seen, the giant pulse intensity ratios vary by about an order of magnitude on short time
scales but follow the smooth variation of the ratios of the main pulse on the longer time scales. Therefore, the former are probably
caused by the specific properties of the giant pulses while the latter are caused by ISS.
This finding allowed us to use the latter variations for correction of the apparent giant pulse flux densities. The giant pulse intensities
in each sideband were multiplied by the factor $\langle I\rangle /I(t)$ where $\langle I\rangle$ is the mean intensity of the normal
main pulse over the whole observing time and $I(t)$ the intensity of the giant pulse at time,~$t$. Such correction reduces the observed
giant pulse intensity to the mean intensity in the upper and lower sidebands.
We are now in a position to analyze the giant pulse intensities and the energy distribution without bias of the ISS.

\subsubsection{Peak Flux Density and Energy of Giant Pulses} \label{peak}

How is the peak intensity or flux density related to the energy of giant pulses? The flux density along the profile of a giant pulse
integrated over a window covering the whole giant pulse duration gives the total spectral energy of the giant pulse received per unit area,
or in short, the energy of the giant pulse.

An accurate computation of the energy is rather difficult, since it depends on the selection of the window for the integration which could
be affected by noise. Although in general the peak giant pulse flux density considerably exceeds the noise level, it drops exponentially
after the peak while the noise level remains constant. At a width of 200~ns, the largest for giant pulses at the 1/e-level (see
Figure~\ref{fig3}), the noise energy roughly equals the energy of the weakest giant pulses caught near the detection threshold. To
minimize the noise contribution and obtain fairly accurate values of the giant pulse energies, we used a variable width of the window for the
integration depending on the current amplitude of the giant pulse,~$A$. This width is given by the width of the exponential decay
function $ A\cdot \exp(-t/\tau)$ at the level of twice the rms noise level and with $\tau = 65$~ns as the scattering time determined above
in~\S\S~\ref{stat},~\ref{wave}.

We found that the energy of giant pulses detected in the main pulse window exceeds by 3~times the mean energy of the main pulse itself.
The energies of the strongest and weakest main pulse giant pulse are 60 and 0.3~times the mean energy of the normal main pulse,
respectively. In the interpulse window the mean giant pulse energy is 5~times the mean energy of the normal interpulse. The difference in
the ratios for the main pulse and the interpulse is however somewhat dependent on the threshold for the detection limit of giant
pulses. The energies of the strongest and weakest interpulse giant pulses are 50 and 0.4~times the mean energy of the interpulse itself.  

\begin{figure}
\includegraphics[width=0.5\textwidth]{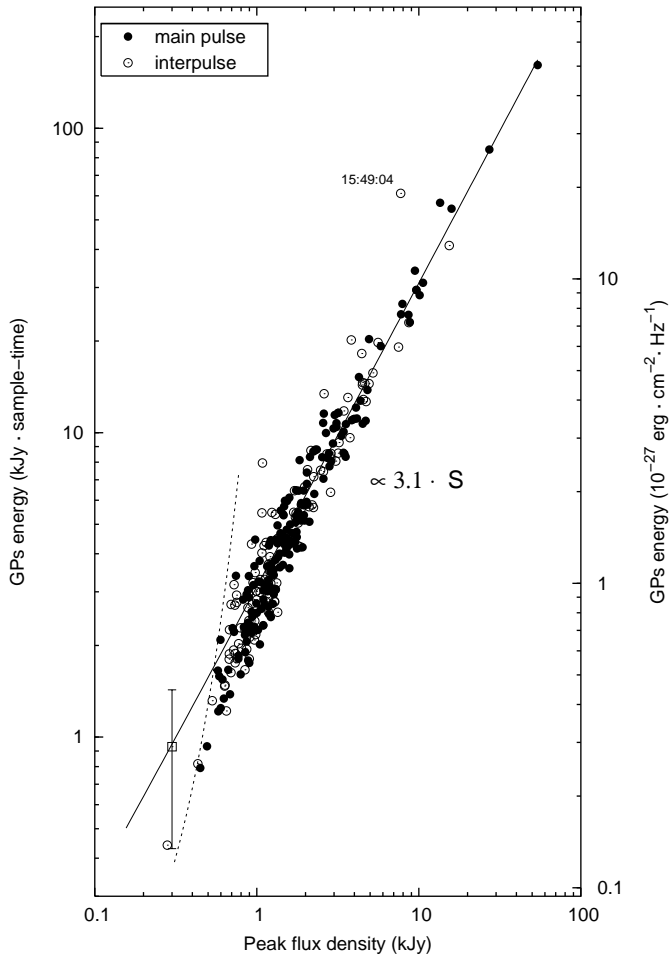}
\caption{
The energy, $\mathrm{E_p}$, of giant pulses (GPs) as a function of their peak flux density, $S$, in the main pulse (filled circles)
and interpulse windows (open circles). The solid line represents a least-squares linear fit to all the data with
$\mathrm{E_p} \propto 3.1\cdot\mathrm{S}$. The energy, $\mathrm{E_p}$, is expressed in units of kJy$~\!\cdot\!$~sample-time
(1~kJy$~\!\cdot\!$~sample-time $ = 3.125\times 10^{-28}$~\mbox{erg~$\cdot$~cm${}^{-2}\!$~Hz${}^{-1}$}). Note that the
slight curvature of the data in the lower left of the Figure is not apparent in a Figure with linear scales on
both axes and does not contribute significantly to the proportionality constant of 3.1. The curved dashed line shows
our noise threshold limit. The bar at the bottom of the plot shows the uncertainty in the low energy data caused by
noise. The same bar at high energies, the upper end of the plot, would appear smaller than the data points.
\label{fig11}
}
\end{figure}

In Figure~\ref{fig11} we plot the peak flux density versus the energy of giant pulses. No significant differences are apparent for main pulse
and interpulse giant pulses. We find a fairly linear relation with a proportionality constant of~3.1 determined from a least-squares fit.
This relation confirms our earlier conclusion that the intrinsic width of giant pulses is indeed very narrow. The giant pulses are broadened
by a constant factor most likely through ISS so that the ratio between the peak flux density and the energy remains also largely constant.
One giant pulse we previously mentioned, at 15:49:04~UTC with a flux density of 8~kJy significantly above the noise, deviates somewhat from
the linear relation. It may be an exception and indicative of intrinsic structure.

\placefigure{fig11}

\subsubsection{Distribution of Giant Pulse Energies} \label{ampldistr}

Having determined the impact of the interstellar medium on several aspects of our observations, we now  unravel properties of the
giant pulses that would otherwise be significantly affected and perhaps even completely masked by scintillation effects.

\begin{figure}
\includegraphics[width=0.5\textwidth]{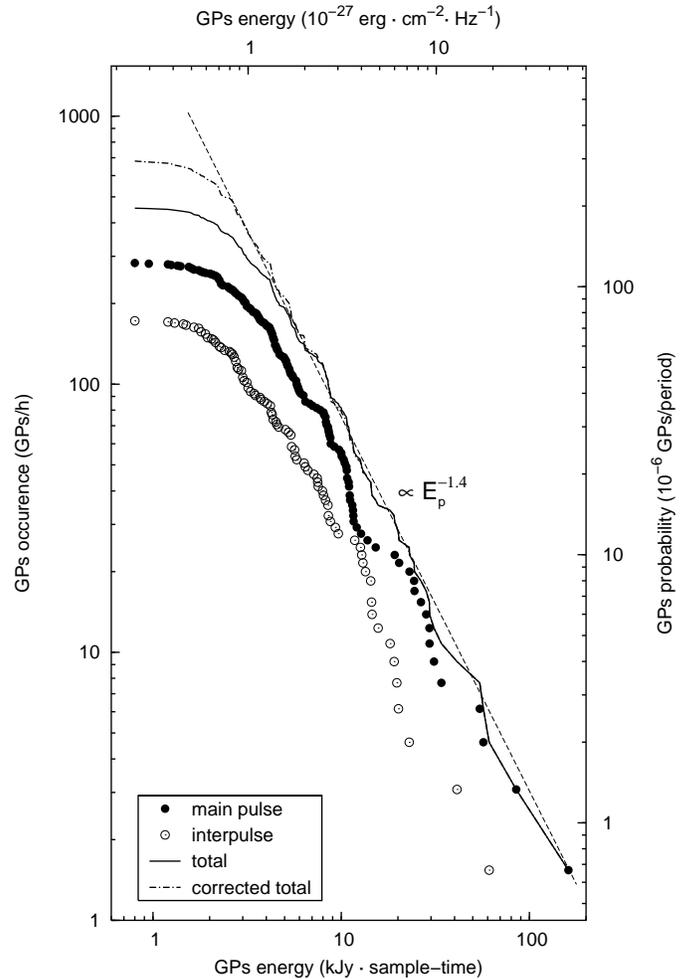}
\caption{
The cumulative distribution of the energies of giant pulses (GPs) in the main pulse (filled circles) and the interpulse windows
(open circles) and of all giant pulses (solid line). The dashed straight line represents a least-squares power law fit to the energies
of all giant pulses with $N \propto \mathrm{E_p}^{-1.4}$ for energies higher than $1.5\cdot 10^{-27}$~\mbox{erg~$\cdot$~cm${}^{-2}\!$~Hz${}^{-1}$}
(1~kJy$~\!\cdot\!$~sample-time $ = 3.125\times 10^{-28}$~\mbox{erg~$\cdot$~cm${}^{-2}\!$~Hz${}^{-1}$}). The gradual
flattening of the curves at low energies is caused by interstellar scintillation. The corrected curve is shown by the dotted-dashed line.
\label{fig12}
}
\end{figure}

In Figure~\ref{fig12} we plot the cumulative distribution of the energies of all giant pulses detected over the whole 39~minutes of
observations and separately for the events in the main pulse  and interpulse windows. No drastic differences are apparent between the
curves for the main pulse and interpulse giant pulses. Above the energy of 4.8~kJy$~\!\cdot\!$~sample-time the cumulative distribution
for all giant pulses is well fit by a power law,
\begin{equation}
N = N_0\cdot\mathrm{E_p}^{\alpha}~,
\label{power}
\end{equation}
with an index of $\alpha=-1.40 \pm 0.01$, where the error is purely formal and where $N$ is the rate of giant pulse occurrence in units of
the number of giant pulses per hour with the giant pulse energy greater than $\mathrm{E_p}$ in kJy$~\!\cdot\!$~sample-time,
and $N_0 = 1860 \pm 40$.

\placefigure{fig12}

If such a steep power law distribution is extrapolated to low energies, giant pulses would produce a significant peak in the mean
profile \citep[see also,][]{kink2000} which is not observed. Therefore, the power law distribution cannot continue to very low
energies. Instead, it must have a cutoff.

Where does this cutoff set in? Our 17$\sigma$-threshold corresponds to 800~Jy of the peak flux density or to an energy of
2.4~kJy$~\!\cdot\!$~sample-time. The ISS can change this magnitude by a factor of 0.5 to~1.8 (see
Figure~\ref{fig10}) washing out this sharp boundary. Therefore, instead of a sharp break at the threshold energy, the
cumulative distribution shows a gradual flattening to the detection threshold energy. Because this effect can mask a possible true
decreasing $N$ at low $\mathrm{E_p}$, it is important to check that there is no deviation from the power law. If one assumes that
the power law distribution~(\ref{power}) is valid below our threshold, the rate of giant-pulse occurrence can be corrected. As shown in
Figure~\ref{fig12}, the corrected curve continues the power law distribution down
to the energy which corresponds exactly to the detection threshold.

\citet{cognard1996} found for the cumulative probability distribution of the intensities of 60~giant pulses, detected in 44~min at
430~MHz, a power law with $\alpha = -1.8 \pm 0.1$ for giant pulses 15~times stronger than the mean. This result appears to be somewhat
different from ours. Part of the difference can perhaps be explained by their relatively small number of giant pulses and the corresponding
small statistics and, very likely, also by the noise contribution from their choice of a wide window for the integration along the
giant pulse profile.

\begin{figure}
\includegraphics[width=0.5\textwidth]{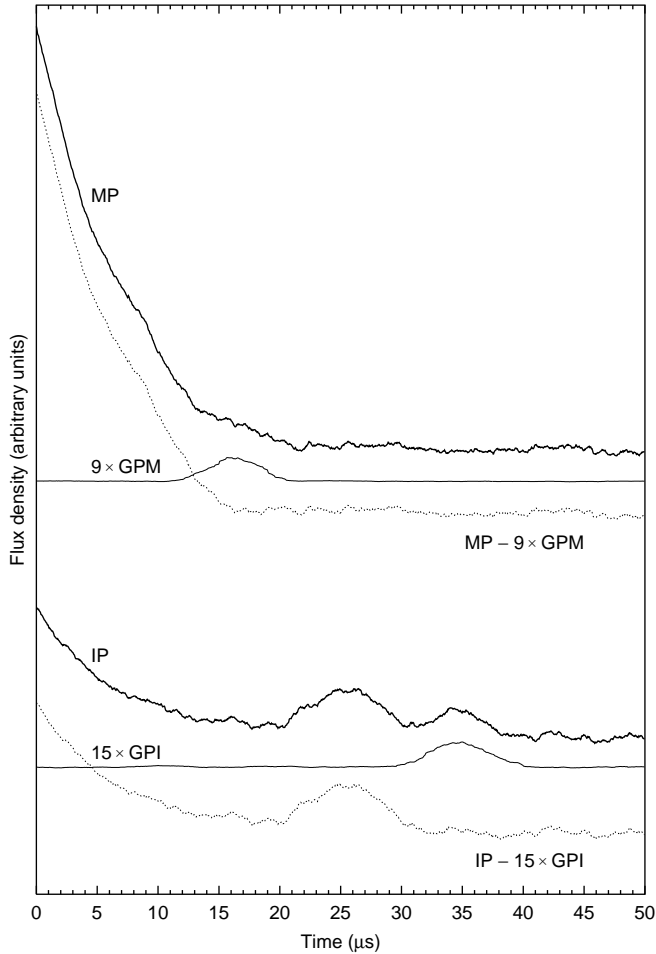}
\caption{
The trailing edges of the mean profile (solid thick line) of the main pulse (MP, upper panel) and the interpulse (IP, lower panel)
together with the corresponding giant pulse profiles. The latter are scaled up by a factor of 9 and~15, respectively (solid thin lines).
Also shown are the difference profiles (dotted curves in the bottom of each panel). The data are smoothed with a $4~\mic$ running interval.
\label{fig13}
}
\end{figure}

\subsubsection{The Cutoff in the Distribution of Giant Pulse Energies
and the Total Rate of Giant Pulses} \label{cutoffdistr}

The cumulative distribution has no cutoff at large energies but does it have a cutoff at low energies, below our threshold? Such a
cutoff can be estimated by taking the average profile of giant pulses and computing how far the power law distribution needs to be
extrapolated toward low energies beyond our threshold to completely account for the average profile. In Figure~\ref{fig13} we show the
trailing edge of the main pulse and interpulse as an enlarged part of Figure~\ref{fig7}. In addition we show the average profiles of
our detected giant pulses both for the main pulse and the interpulse windows (GPM and GPI). Each of the profiles is smoothed to $4~\mic$,
or to about the width of the giant pulse windows.

\placefigure{fig13}

The contribution of the giant pulses can be clearly seen in the average main pulse and interpulse. In the average main pulse the
giant pulse feature is partially blended with the very end of the trailing edge of the average main pulse. In the average interpulse
region the giant pulse feature occurs just after a similarly weak interpulse component but is clearly outside of any regular emission.
However, we had to multiply the average main pulse giant pulse profile by a factor 9 and the average interpulse giant pulse profile by a
factor~15 to account for the equivalent features in the average main pulse and interpulse profiles. Only then were the giant pulse
features subtracted completely in the difference profiles (dotted lines in Figure~\ref{fig13}). In other words, giant pulses clearly also
occurred below our threshold. If we assume that the power law distribution can be extrapolated to low energies beyond our threshold, then
a cutoff in the distributions can be calculated. This cutoff occurs for the main pulse giant pulses at a minimum flux density of 16~Jy,
corresponding to an energy of~$2.125\times 10^{-29}$~\mbox{erg~$\cdot$~cm${}^{-2}\!$~Hz${}^{-1}$} and for the
interpulse giant pulses at a minimum flux density of 5~Jy, corresponding to an energy of~$1.0625\times
10^{-29}$~\mbox{erg~$\cdot$~cm${}^{-2}\!$~Hz${}^{-1}$}. With this cutoff, the total number of the giant pulses in our 39~minutes
observing time is~$\sim 200\,000$, with~$\sim 50\,000$ in the main pulse and~$\sim 150\,000$ in the interpulse windows. These numbers can
be multiplied by the directivity factors for the main pulse and interpulse giant pulses to yield the total rate
of giant pulses per neutron star revolution. We calculate the directivity factors from the longitude range in which 97\% of our
giant pulses occurred, 1$\fdg$3 for the main pulse giant pulses and 1$\fdg$9 for the interpulse giant pulses (see
above in \S~\ref{arrival}). We obtain directivity factors of the order of 270 and 190 for the main pulse and interpulse giant pulses
respectively. These figures yield a total rate of~$\sim 25$ giant pulses per neutron star revolution. Therefore giant pulses occur
frequently although they are only rarely detected.

\section{DISCUSSION} \label{discussion}

Thirty-nine minutes of observations of the millisecond pulsar B1937+21 with more than tenfold higher time resolution than ever before
have revealed new insight in the extraordinary phenomenon of giant pulses. They are in most cases as narrow or even
narrower than 15~ns and can have flux densities as high as 65\,000~Jy. No upper cutoff is visible in the cumulative distribution of their
energies, and therefore giant pulses with even higher energies and probably also higher flux densities may be detectable in the future. A
lower cutoff, however, exists and allowed us to compute a relatively large number for the frequency of occurrence of giant pulses,
namely of~25 per neutron star revolution.

Further, as pointed out by others before and confirmed here in more detail, giant pulses
occur in a very narrow longitudinal range, $\tau_\mathrm{\scriptscriptstyle{WGP}}$, of less than $10~\mic$ at the
very trailing edge of the regular profile. The narrow range means that the emission is
confined to within a beam of~$\theta$:
\begin{displaymath}
\theta <
\frac{2\pi\cdot\tau_\mathrm{\scriptscriptstyle{WGP}}}{\mathrm{P}}\sim
2\arcdeg~,
\end{displaymath}
where $\mathrm{P} = 1.557$~ms is the pulsar's period.

These characteristics of giant pulses lead to important estimates of their physical parameters and to speculations as to their physical
nature. If giant pulses are temporal fluctuations within their confined longitudinal range rather than intensity variations due to
the sweep of a longitudinally modulated beam, then the diameter,~$d$, of the sources from which they originate have to be extremely small,
namely $d < c\tau_\mathrm{\scriptscriptstyle{GP}}$, or less than 15~light nanoseconds or 4.5~m. If so, then these are the smallest
emitters found in the universe apart from those found for the Crab pulsar with a size even an order of magnitude smaller \citep{hankins2003}.

Their brightness temperature is enormous. It can be estimated as
\begin{displaymath}
T_\mathrm{b} =
\frac{1}{k}~\frac{\mathrm{E_p}}{\tau_\mathrm{\scriptscriptstyle{GP}}}{\biggl(\frac{c}{\nu}\biggr)}^2{\left(\frac{L}{d}\right)}^2
> \frac{\mathrm{E_p}\cdot
L^2}{k~\nu^2\tau_\mathrm{\scriptscriptstyle{GP}}^3}~,
\end{displaymath}
where $k = 1.38 \cdot 10^{-16}$~\mbox{erg~$\cdot$~K${}^{-1}$}~is Boltzmann's constant, $\mathrm{E_p}$~the energy of the giant pulse,
$L$~the distance to the pulsar of~3.6~kpc \citep{taylor1993}, $\tau_\mathrm{\scriptscriptstyle{GP}}$~the width of the giant pulse,
and $\nu$~the observing frequency. For our strongest, 65\,000~Jy, giant pulse with $\tau_\mathrm{\scriptscriptstyle{GP}} \leq 15$~ns and
$\mathrm{E_p} = 5\cdot 10^{-26}$~\mbox{erg~$\cdot$~cm${}^{-2}\!$~Hz${}^{-1}$} (see Figure~\ref{fig12}), the brightness temperature is
$T_\mathrm{b} \geq 5\cdot 10^{39}$~K. This is the highest brightness temperature ever detected in the universe.
The weakest giant pulse at the low-end cutoff of the cumulative energy distribution with
$\mathrm{E_p} = 10^{-29}$~\mbox{erg~$\cdot$~cm${}^{-2}\!$~Hz${}^{-1}$} still has $T_\mathrm{b} > 10^{36}$~K.

How do the charactersitics of the giant pulses of B1937+21 compare with those of giant pulses of other pulsars? However, although giant
pulses were reported for a few other pulsars, the only other giant pulses with well studied characteristics are those from the Crab
pulsar. We therefore compare the characteristics of the giant pulses from these two pulsars.

The giant pulses from the millisecond pulsar B1937+21 are very short. The ones from the Crab pulsar also contains extremely short
elements with a width $\leq 2$~ns \citep{hankins2003}. Their brightness temperature reaches $10^{38}$~K almost comparable to
that of the giant pulses of B1937+21. Further, they are nearly 100\%~circularly polarized either with right- or left-handed
orientation, just as observed for the giant pulses of B1937+21~\citep{cognard1996, popov2004}. In addition, they occupy a
relatively narrow window of~$3\degr$ for the Crab pulsar and~$1\degr$ for B1937+21. Also, both pulsars have a power law cummulative
distribution of giant pulse energies with low-energy cutoffs but no high-energy cutoffs.

The main difference appears to be that the relatively narrow windows of giant pulses from the Crab pulsar cover the main pulse and
interpulse region whereas in B1937+21 they cover just the very trailing edge of the main pulse and are located even slightly outside
the interpulse region.

What is the physical nature of the giant pulses? In general, giant pulses are believed to be a specific feature of the pulsar's radio
emission which still can be explained in terms of the general emission mechanism working inside the neutron star magnetosphere, as a
result of nonlinear interactions of waves and particles in ultrarelativistic plasma
\citep{petrova2004, asseo1998, weatherall1998, onishchenko1990}.

A somewhat different  explanation of the origin of giant pulses was suggested by~\citet{istomin2003}. He suggested that giant pulses are
the result of the reconnection of the outermost open magnetic field lines from the two poles of the pulsar
in the light cylinder region. Such giant pulses must originate close to the last closed magnetic field
line, i.e. on the very edge of the average pulse profile. This is indeed observed for the millisecond pulsar B1937+21 although not
in the Crab pulsar.

The most decisive clues as to the nature of the giant pulses may come from considering the enormous radiation energy densities of giant
pulses. As the relatively large radiation energy density of micropulses had largely ruled out relativistic beaming at the light
cylinder as an interpretation for such pulses some 25 years ago \citep[\eg,][]{manchester1973, bartel1978, bartel1982},
the radiation energy density could constrain interpretations of the giant pulses.

In case of the Crab pulsar the nanopulses's radiation energy density is very close to the plasma energy density \citep{hankins2003}.
These authors suggested that giant pulses are the result of the focusing of many plasma waves in a small volume.

However, B1937+21 has a magnetic field at its surface which is 1000~times weaker than that of the Crab pulsar. How does in this case
the radiation energy density compare with the plasma energy density and the magnetic field energy density at different heights from the
surface of the neutron star?

The volume density of the radiation energy is given as
\begin{displaymath}
u_\mathrm{\scriptscriptstyle{r}}^\mathrm{\scriptscriptstyle{GP}} \simeq
\frac{\mathrm{E_p}\Delta\nu}{W\tau_\mathrm{\scriptscriptstyle{GP}} c}~,
\end{displaymath}
where $W$ is a dilution factor with
\begin{displaymath}
W = {\left(\frac{d}{\theta L}\right)}^2 <
{\left(\frac{\mathrm{P}~\tau_\mathrm{\scriptscriptstyle{GP}}
c}{2\pi\tau_\mathrm{\scriptscriptstyle{WGP}}L}\right)}^2
\end{displaymath}
and $\Delta\nu$ is the bandwidth of the giant pulse. Since the latter has to be at least as wide as $\Delta\nu
\sim\tau_\mathrm{\scriptscriptstyle{GP}}^\mathrm{\scriptscriptstyle{-1}}$, we finally have
\begin{displaymath}
u_\mathrm{\scriptscriptstyle{r}}^\mathrm{\scriptscriptstyle{GP}} \gtrsim
\mathrm{E_p}\cdot{\left(\frac{2\pi
L}{\mathrm{P}}\right)}^2\frac{\tau^2_\mathrm{\scriptscriptstyle{WGP}}}{c^3\tau^4_\mathrm{\scriptscriptstyle{GP}}}~.
\end{displaymath}

The upper limit for $\tau_\mathrm{\scriptscriptstyle{GP}} \leq 15$~ns gives the corresponding lower limit for
$u_\mathrm{\scriptscriptstyle{r}}^\mathrm{\scriptscriptstyle{GP}} \geq 7\cdot 10^{15}$~erg~$\cdot$~cm${}^{-3}$ for the strongest detected
giant pulse with $\mathrm{E_p} = 5\cdot 10^{-26}$~erg~$\cdot$~cm${}^{-2}\!$~Hz${}^{-1}$ (see Figure~\ref{fig12}) and
$u_\mathrm{\scriptscriptstyle{r}}^\mathrm{\scriptscriptstyle{GP}} > 2\cdot 10^{12}$~erg~$\cdot$~cm${}^{-3}$ for the possible weakest
one at the low-end cutoff where $\mathrm{E_p} = 10^{-29}$~erg~$\cdot$~cm${}^{-2}\!$~Hz${}^{-1}$~.

It is generally assumed that near the light cylinder equipartition prevails between the energy density of the magnetic field,
$u_\mathrm{\scriptscriptstyle{B}}$, and that of the relativistic plasma, $u_\mathrm{\scriptscriptstyle{p}}$. It can be estimated as
$u_\mathrm{\scriptscriptstyle{B}}^\mathrm{\scriptscriptstyle{LC}}\sim u_\mathrm{\scriptscriptstyle{p}}^\mathrm{\scriptscriptstyle{LC}}\sim 4
\cdot 10^{10}$~erg~$\cdot$~cm${}^{-3}$, if we assume a neutron star radius $r_\star = 10^6$~cm,
a moment of inertia $I = 10^{45}$~g~$\cdot$~cm${}^2$, and take for the pulsar period $\mathrm{P} = 1.557$~ms and for the
period derivative $\dot{\mathrm{P}} = 1\cdot 10^{-19}$~s/s \citep[see][]{kaspi1994}.
Near the neutron star's surface we get instead
$u_\mathrm{\scriptscriptstyle{p}}^\mathrm{\scriptscriptstyle{NS}}\sim 2\cdot 10^{13}$~erg~$\cdot$~cm${}^{-3}$,
$u_\mathrm{\scriptscriptstyle{B}}^\mathrm{\scriptscriptstyle{NS}}\sim 7\cdot 10^{15}$~erg~$\cdot$~cm${}^{-3}$. In other words, the
radiation energy density of giant pulses from B1937+21 exceeds by orders of magnitude the plasma energy density everywhere inside the
magnetosphere. Even at the neutron star's surface the radiation energy density of our strongest pulse is more than 300~times larger
than the computed plasma energy density at that location. Hence, any plasma mechanism is unlikely as an interpretation of the giant pulses.
Moreover, for the strongest giant pulses the radiation energy density is comparable to, or even larger than, the magnetic field energy
density near the neutron star's surface. However inside the pulsar magnetosphere the radiation energy density must be smaller than the
magnetic field energy density and can only be equal to it as an absolute limit for any theoretically conceivable physical processes
inside this magnetosphere. However in reality it appears to be almost impossible to find a physical mechanism which is able to provide
almost complete conversion of the magnetic field energy into radio emission energy.

It appears that the nature of the giant pulses has to be found elsewhere. We suggest that giant pulses are the direct observable
result of the polar gap discharge, the basic process generating energetic particles and exciting the magnetosphere. A volume charge
created under the polar cap at the first stage of gap discharge may be considered as a natural cause of giant pulse emission.  

In this context the recent RXTE X-ray observations of B1937+21 by \citet{cusumano2003} are of particular interest. It was found that the
maxima of the X-ray profiles, both, of the main pulse and the interpulse, coincide with the respective giant pulse windows. This is
evidence for giant pulses being indeed related to the high energy processes of the polar gap discharge. Therefore, giant pulses may provide
us with an opportunity to directly study the main physical process responsible for the pulsar radiation.
 
\section{CONCLUSIONS} \label{conclusions}

Here we give our conclusions of our findings.

\begin{enumerate}
\item Giant pulses of B1937+21 are extremely short, generally $\leq 15$~ns. These are the shortest pulses found so far in any pulsar
after those of the Crab pulsar.

\item The strongest giant pulse had a flux density of 65\,000~Jy. Coupled with a width of $\leq 15$~ns, the brightness temperature is
$T_\mathrm{b} \geq 5\cdot 10^{39}$~K, the highest observed in the universe.

\item Giant pulses occur in general with a single spike. Only in one case did we find a more complex giant pulse.

\item The cummulative distribution of energies of giant pulses is a power law with an index of $-1.4$. No high energy cutoff could be
found. However, we inferred a low energy cutoff corresponding to a flux density of~$\sim 10$~Jy. Correspondingly we estimate the total
number of detected and undetected giant pulses to be~$\sim 200\,000$ in our data from 39~min of observations or, corrected for the
directivity factor,~$\sim 25$ giant pulses per neutron star revolution. A giant pulse with a flux density of 65\,000~Jy is expected to be
observed only once per hour.

\item Giant pulses occur in a very narrow range of longitudes, the vast majority in a $5.8~\mic$ and $8.2~\mic$ window at the very
trailing edges of the regular main pulse and interpulse profiles, respectively. Numerically, the centers of these windows are
delayed by $58.3 \pm 0.3$ and $65.2 \pm 0.5~\mic$ from the peaks of the average regular main pulse and interpulse profiles,
correspondingly.

\item Although giant pulses are found close in longitude to the regular main pulses and interpulses, no relation in their emission
intensities was found. Also, no relation between the flux densities of the giant pulses in the main pulse and interpulse windows was
observed.
 
\item The radiation energy density of the strongest giant pulses we detected is more than 300~times larger than the computed plasma
energy density at the surface of the neutron star and even higher than the magnetic filed energy density at that surface. Therefore
any plasma mechanism is unlikely to be able to account for the giant pulses.

\item We suggest that giant pulses are directly related to the discharges happening in the polar cap region.
\end{enumerate}

\acknowledgements

This investigation was supported by the Russian Foundation for Fundamental Research (project number 04-02-16384). Research at York
University is supported by grants from the Canadian NSERC. Observations were carried out at the DSN \mbox{70-m} telescope operated
by JPL/Caltech under contract with the National Aeronautics and Space Administration.

\bibliographystyle{apj}
\bibliography{biblio}

\end{document}